\documentclass[aps,prl,superscriptaddress,showkeys,showpacs,twocolumn,longbibliography,nofootinbib]{revtex4-1}
\usepackage{amsmath}
\usepackage{float}
\usepackage[normalem]{ulem}
\usepackage{braket}
\usepackage{graphicx}
\usepackage{enumitem}
\usepackage{multirow}
\usepackage[colorlinks=true, pdfstartview=FitV, linkcolor=red, citecolor=blue, urlcolor=blue]{hyperref}
\usepackage{slashed}

\begin{document}
\title{Bayesian Inference of the Magnetic Component in the Quark-Gluon Plasma}

\author{Yu Guo}  
\address{Department of Physics, Tsinghua University, Beijing 100084, China. }

\author{Jinfeng Liao} \email{liaoji@iu.edu}
\address{Physics Department and Center for Exploration of Energy and Matter, Indiana University, 2401 N Milo B. Sampson Lane, Bloomington, IN 47408, USA.}

\author{Shuzhe Shi} \email{shuzhe-shi@tsinghua.edu.cn}
\address{Department of Physics, Tsinghua University, Beijing 100084, China. }
\affiliation{State Key Laboratory of Low-Dimensional Quantum Physics, Tsinghua University, Beijing 100084, China.}

\date{\today}

\begin{abstract}
    The chromo-magnetic monopoles (CMM), emergent topological excitations of non-Abelian gauge fields carrying chromo-magnetic charge, have long been postulated to play an important role in the vacuum confinement of quantum chromodynamics (QCD), the deconfinement transition at temperature $T_c\approx 160\rm MeV$, as well as the strongly coupled nature of quark-gluon plasma (QGP). While such CMMs were found to help provide resolutions for challenging puzzles from heavy-ion collision measurements, they were typically introduced as model assumptions in the past. Here we perform a data-driven Bayesian inference of the chromo-magnetic component, within the CUJET framework, and extract its temperature-dependent abundance for the first time. Using the \textsc{cujet3} framework for calculations of jet energy loss and analyzing a comprehensive experimental data set for nuclear modification factor ($R_{\mathrm{AA}}$) and elliptic flow ($v_2$) of high-transverse-momentum hadrons, the fraction of CMMs in the QGP is obtained by Bayesian inference and is found to be substantial in the $1\sim 2 T_c$ region. The  posterior CMM fraction is further tested against additional data and is also shown to predict QGP transport properties that are quantitatively consistent with the state-of-the-art knowledge.
\end{abstract}

\maketitle

\emph{Introduction.}---
Fifty years after its discovery as the fundamental theory of  strong interaction, quantum chromodynamics (QCD) remains challenging to decipher. While QCD is based on quarks and gluons, they are forced to hide deeply inside hadrons. Such an outstanding puzzle of vacuum confinement has so far defied an explanation~\cite{tHooft:1999cgx,Greensite:2011zz,Gross:2022hyw}, although it has long been postulated that chromo-magnetic monopoles (CMM) play a key role. These CMMs are emergent topological configurations of non-Abelian gauge fields that carry chromo-magnetic charge and source long-range magnetic fields~\cite{tHooft:1974kcl,Polyakov:1974ek}.  They become localized and abundant at strong gauge coupling. In the case of QCD vacuum, the CMMs 
 are proposed  to form a condensation, thus making the QCD vacuum a 't Hooft--Mandelstam ``dual superconductor'' , where  chromo-electric quarks and gluons are confined inside color-neutral hadrons~\cite{Mandelstam:1974pi,tHooft:1977nqb,tHooft:1981bkw}.

\begin{figure*}
    \centering
    \includegraphics[width=0.8\textwidth]{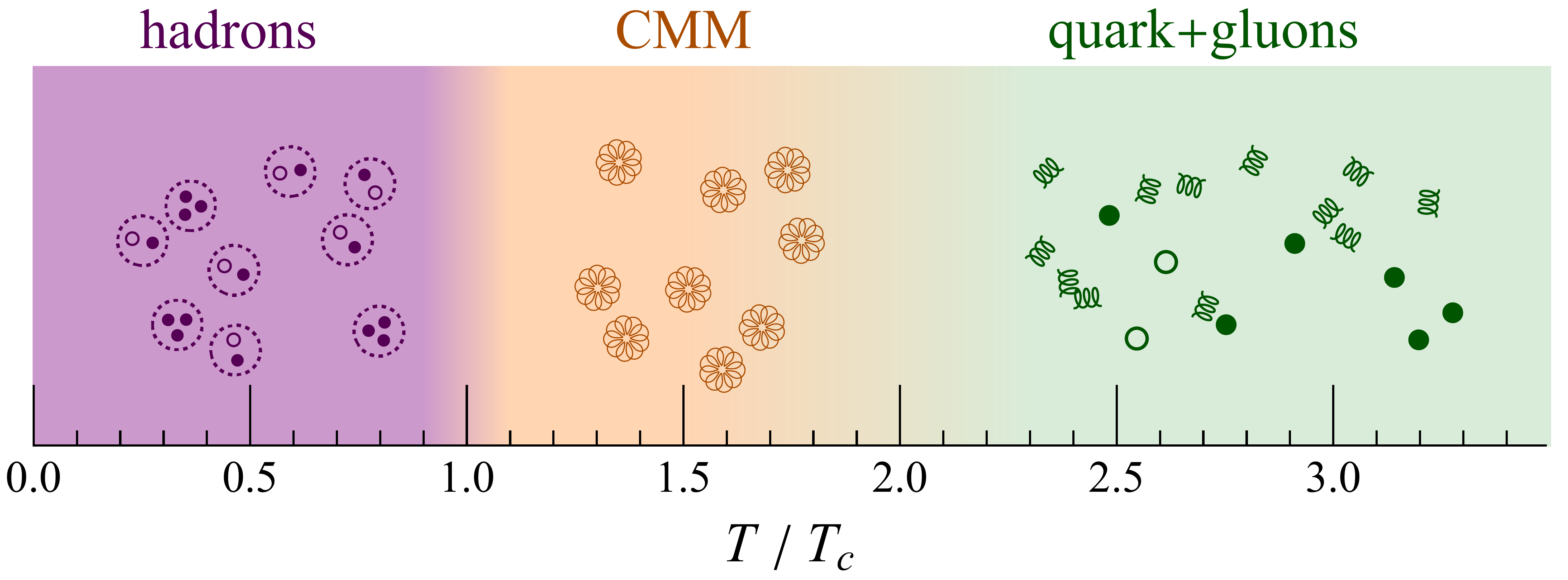}
    \caption{An illustration of the dominant microscopic  degrees of freedom in QCD matter at different temperatures. At low temperature ($T<T_c$), quarks and gluons are confined into a gas of mesons and baryons. At  high temperature ($T\gtrsim 2T_c$), quarks and gluons are fully liberated and dominant, gradually approaching a weakly coupled quark-gluon plasma.   An intermediate phase emerges in the $1\sim 2T_c$ region where the chromo-magnetic monopoles are the dominant degrees of freedom and lead to properties of a strongly coupled fluid as observed in heavy ion collision experiments.
    \label{fig:illustration}}
\end{figure*}

With increasing temperatures, such a condensate of CMMs is expected to ``evaporate'' eventually and lead to the deconfinement transition at  $T_c\approx 160\rm MeV$. 
At extremely high temperatures where QCD coupling becomes small due to asymptotic freedom, one expects a weakly coupled quark-gluon plasma (QGP). However, for temperatures just above $T_c$, the QCD coupling is still strong and one expects the CMMs originally in the vacuum condensate to become thermal degrees of freedom. An intermediate phase of a strongly coupled QGP in $1\sim 2 T_c$ region has been proposed, in which a magnetic component of abundant CMMs is dominant whiles the chromo-electric quarks and gluons are  only partially liberated~\cite{Liao:2005pa, Liao:2006ry, Liao:2008jg, Liao:2008dk, Chernodub:2006gu,DAlessandro:2007lae,Bonati:2013bga, Hidaka:2008dr,Hidaka:2009ma,Shuryak:2008eq,Ratti:2008jz,Ramamurti:2017zjn,Fujimoto:2025sxx}. This scenario is illustrated in Fig.~\ref{fig:illustration}. Clearly, a quantitative characterization of such a magnetic component will significantly deepen our understanding of QCD confinement, its deconfinement transition, and the properties of a strongly coupled QGP.

Relativistic heavy-ion collision experiments performed  at BNL's Relativistic Heavy-Ion Collider (RHIC) and CERN's Large Hadron Collider (LHC) over the past decades have provided unique opportunities for systematic studies of the QGP. 
Noticeably, measurements from RHIC and LHC have enabled empirical constraints on key transport properties of QGP, such as the specific shear viscosity, and established QGP as a strongly coupled plasma with nearly perfect fluidity.

A natural question is what one can learn about the QGP magnetic component by confronting its phenomenological consequences with  experimental data. One distinctive example is about the  long-standing ``$R_\mathrm{AA} \otimes v_2$ puzzle'' in the jet quenching phenomenon.  Energetic quarks and gluons created from initial hard scatterings, known as jets, are penetrating probes of the QGP~\cite{Wang:1992qdg, Gyulassy:2000gk}. They scatter with medium constituents and lose part of their initial energy via both elastic and inelastic collisions  before turning into hadrons via fragmentation. This energy loss is manifested  as a reduced yield of high transverse momentum ($p_T$) hadrons compared to baseline expectations from independent proton-proton collisions. 
The nuclear modification factor $R_{\mathrm{AA}}$ quantifies this suppression by taking the ratio between hadron differential spectrum in nucleus-nucleus (AA) collisions and that in proton-proton ($pp$) collisions scaled by the number of initial binary collisions ($N_\mathrm{bin}$),
\begin{align}
\begin{split}
&R_\mathrm{AA}(p_T, \phi) 
\\\equiv\;& 
    \frac{\mathrm{d}^3 N_\mathrm{AA}(p_T, \phi)/p_T\mathrm{d}p_T\, \mathrm{d}\phi\, \mathrm{d}\eta}{N_\mathrm{bin}\, \mathrm{d}^3 N_{pp}(p_T, \phi)/p_T\mathrm{d}p_T\, \mathrm{d}\phi\, \mathrm{d}\eta}
\\=\;&
    R_\mathrm{AA}(p_T) \Big(1 + 2v_2(p_T)\cos(2\phi-2\Psi_{2}) 
    %\\&
    %+ 2v_3(p_T)\cos(3\phi-3\Psi_{3}) 
    + \cdots \Big)\,.
\end{split}
\end{align} 
In the above, its dependence on azimuthal angle ($\phi$) is mainly captured by an elliptic harmonic coefficient $v_2$.
As it turns out, a simultaneous description of both $R_{AA}$ and $v_2$ with the same parameter set has been a challenge for various jet quenching models with different assumptions of medium constituents and the exact energy-loss mechanism, including \textsc{amy}~\cite{Arnold:2001ba, Arnold:2001ms, Arnold:2002ja}, 
\textsc{asw}~\cite{Wiedemann:2000za, Salgado:2003gb, Armesto:2003jh, Armesto:2004pt, Armesto:2005iq}, \textsc{dglv}~\cite{Gyulassy:1993hr, Gyulassy:2000er, Djordjevic:2003zk, Djordjevic:2008iz,Buzzatti:2011vt, Xu:2014ica}, \textsc{ads-cft}~\cite{Casalderrey-Solana:2014bpa}, \textsc{lbt}~\cite{He:2015pra, Cao:2016gvr}, higher-twist~\cite{Guo:2000nz, Wang:2001ifa, Majumder:2007ae}, geometric models~\cite{Shuryak:2001me, Drees:2003zh, Betz:2014cza, Noronha-Hostler:2016eow}, etc.  
It was then shown that by introducing the magnetic component of QGP, the jet-medium interaction becomes significantly enhanced in the near-$T_c$ temperature region, due to the fact that the coupling of a quark or gluon jet with a medium monopole is ``dictated'' by the famous Dirac quantization condition $\alpha_E \, \alpha_M =1$~\cite{Dirac:1931kp, Liao:2008jg, Liao:2008dk}. Such near-$T_c$ enhancement due to CMM results in a stronger elliptic anisotropy in jet energy loss and has been demonstrated by quantitative modeling, notably via the \textsc{cujet3} framework, to resolve this puzzle~\cite{Xu:2014tda, Xu:2015bbz, Shi:2018lsf, Shi:2018izg}.
Other recently  attempted solutions also indicate the need for enhanced near-$T_c$ contributions: either through increased energy loss at lower temperatures~\cite{Andres:2019eus} or by involving soft partons near freeze-out hypersurface via coalescence~\cite{Zhao:2021vmu}.

Despite the phenomenological success of the magnetic component in QGP, past studies such as \textsc{cujet3.1} suffer from the fact that the existence and abundance of CMM are model assumptions, estimated from the suppression of chromo-electric components -- either via the Polyakov loop~\cite{Hidaka:2008dr, Hidaka:2009ma, Dumitru:2010mj, Lin:2013efa} or quark number susceptibility~\cite{McLerran:1987pz, Gottlieb:1988cq, Gavai:1989ce, Gottlieb:1987ac}. 
The goal of this work is to provide the first statistically rigorous, data-driven extraction of the CMM fraction in QGP. As we shall show, by using the \textsc{cujet3} framework for calculating jet energy loss and analyzing a comprehensive experimental data set for  $R_{\mathrm{AA}}$ and $v_2$ of high $p_T$ hadrons, the fraction of CMMs in the QGP is obtained by Bayesian inference and is found to be substantial in the $1\sim 2 T_c$ region.

\vspace{3mm}
\emph{The \textsc{cujet} framework.}---
The \textsc{cujet} framework describes the evolution of initial hard partons through a hot QCD plasma with hydrodynamic expansion, taking into account the in-medium energy loss via both elastic collisions and inelastic gluon radiation.
The former is implemented according to  the Thoma--Gyulassy (TG) elastic energy loss formula \cite{Thoma:1990fm, Bjorken:1982tu, Peigne:2008nd} with the Peign\'{e}--Peshier running coupling prescription~\cite{Peigne:2008nd}: see~\cite{Xu:2014ica, Xu:2014tda, Xu:2015bbz, Shi:2018izg} for details.  
The latter, which is the dominant energy loss mechanism especially for light partons, is described by the \textsc{dglv} formalism~\cite{Gyulassy:1999zd, Gyulassy:2000er, Djordjevic:2003zk, Djordjevic:2008iz}.  To leading order in opacity expansion,  the  probability density of emitting a gluon by a jet parton of energy $E$  with energy fraction $x_E$, that incorporates the Landau--Pomeranchuk--Migdal effect, is given by: 
\begin{align}
\begin{split}
&x_E \frac{dN^{n=1}_g}{dx_E} 
\\=\,&
    \frac{3C_R}{\pi^2} \int d\tau\,\Gamma \int d^2 \mathbf{k}_\perp\, \alpha_s\Big(\frac{\mathbf{k}_\perp^2}{x_+(1-x_+)}\Big)
\\& 
    \int d^2 \mathbf{q}_\perp\left(
    (3\rho_g + \frac{4}{3}\rho_q)
    \frac{\mathrm{d}\sigma_{E}}{\mathrm{d}\mathbf{q}_\perp^2}
+
    3\rho_m\frac{\mathrm{d}\sigma_{M}}{\mathrm{d}\mathbf{q}_\perp^2}
    \right)
\\&
    \left(
    \frac{\mathbf{k}_\perp - \mathbf{q}_\perp}{(\mathbf{k}_\perp - \mathbf{q}_\perp)^2 + \chi^2} \cdot
    \Big( 
    \frac{ \mathbf{k}_\perp - \mathbf{q}_\perp }{(\mathbf{k}_\perp - \mathbf{q}_\perp)^2 + \chi^2} 
    -
    \frac{\mathbf{k}_\perp}{\mathbf{k}_\perp^2 + \chi^2} 
    \Big) 
    \right) 
\\&
    \left(
    1 - \cos\Big( \frac{ (\mathbf{k}_\perp - \mathbf{q}_\perp)^2 + \chi^2 }{2x_+E}\tau \Big) 
    \right)
    \;\left| \frac{dx_+}{dx_E} \right|
\,.
\end{split}\label{eq:dN/dx}
\end{align}
In the above, $C_R$ is the quadratic Casimir factor being $3$ (or $4/3$) for a gluon (or quark) jet, $\Gamma=n^\mu_\mathrm{jet} u_\mu$ is the Lorentz contraction factor, and $\chi^2 = M^2 x_+^2 + (1-x_+)f_E^2 \mu^2/2$ regularizes the collinear divergences with $x_+ = (1 + \sqrt{1-(k_\perp/x_E E)^2})x_E/2$ the fractional plus-momentum. 
The cross sections for jet-medium scattering   are given by 
\begin{align}
    \frac{\mathrm{d}\sigma_{E}}{\mathrm{d}\mathbf{q}_\perp^2} = \frac{f_E^2\, \alpha_s^2(\mathbf{q}_\perp^2)}{(\mathbf{q}_\perp^2 + f_E^2 \mu^2) \, \mathbf{q}_\perp^2}\,,\quad
    \frac{\mathrm{d}\sigma_{M}}{\mathrm{d}\mathbf{q}_\perp^2} = \frac{f_M^2 }{(\mathbf{q}_\perp^2 + f_M^2 \mu^2) \, \mathbf{q}_\perp^2}\,, \nonumber 
\end{align}
with a chromo-electric and chromo-magnetic particle from the medium, respectively. $\rho_g \equiv \chi_T \frac{16\,\rho }{16 + 9 N_f}$, $\rho_q \equiv \chi_T \frac{9 N_f \,\rho}{16 + 9 N_f}$, and $\rho_m = (1-\chi_T)\rho$ are respectively the number densities of gluons, quarks, and chromo-magnetic monopoles, where $\rho$ is the total number density of in-medium quasiparticles. We use $N_f=2.5$ as an effective number of flavors.
Jet-medium interactions are modulated by electric and magnetic screening effects, with the Debye screening scale $\mu(T)$ determined from a self-consistent equation $\mu = \sqrt{4\pi(1+N_f/6)\alpha_s(\mu^2)}\,T$. The additional electric and magnetic screening coefficients are defined, respectively, as $f_E = \sqrt{\chi_T(T)}$ and $f_M = c_M \sqrt{4\pi \alpha_s(\mu)}$, where $c_M$ is a non-perturbative dimensionless magnetic screening coefficient. 
The running coupling $\alpha_s(Q^2)$ is assumed to take an infrared-saturated non-perturbative form:
\begin{equation}
\alpha_s(Q^2) = \frac{\alpha_C}{1 + \frac{9 \alpha_C}{4 \pi} \log(Q^2/\Lambda^2)},
\end{equation}
where $\alpha_C$ is a model parameter that controls the infrared saturation of the coupling strength, and $\Lambda=200~\mathrm{MeV}$ is the QCD non-perturbative scale. The temperature and fluid velocity profiles are computed from \textsc{vishnu2+1} simulation~\cite{Shen:2014vra}. 
In the present implementation, 
energetic partons traverse the evolving medium along eikonal, straight-line trajectories, assuming transverse deflections strongly suppressed by their large longitudinal momenta. 
Transverse momentum transfers enter the elastic and radiative energy-loss kernels, but stochastic deflections of the jet trajectory due to accumulated transverse momentum broadening are not explicitly simulated. Recent studies~\cite{Antiporda:2021hpk, Bahder:2024jpa} suggest that such broadening may generate additional azimuthal modifications and affect $v_2$. See~\cite{Shi:2018lsf,Shi:2018izg} for further details of the \textsc{cujet} framework.

The key element of \textsc{cujet} is the function $\chi_T(T) \in [0,1]$, which quantifies the temperature-dependent chromo-electric fraction of the QCD medium. Obviously, $(1-\chi_T)$ gives the chromo-magnetic fraction. 
 When $\chi_T \to 1$, the QGP is a pure chromo-electric plasma; in the other limit of $\chi_T \to 0$, it is a non-perturbative plasma dominated by chromo-magnetic monopoles. In past studies, $\chi_T$ was implemented with informed model assumptions. Here, we take a completely different approach: instead of assuming the existence and abundance of chromo-magnetic monopoles, we leave $\chi_T(T)$ as free parameters and utilize Bayesian inference to reach a data-driven determination of $\chi_T(T)$, along with the coupling and screening parameters $\alpha_C$ and $c_M$. 
In this framework, the CMM component is implemented as a structural ingredient quantified by $\chi_T(T)$, whose preferred functional form is to be extracted via Bayesian inference from experimental data. Nevertheless, it shall be emphasized that the physical parameter space does include the possibility of arbitrary $T$-dependence as well as of the extreme case with zero or negligible CMM fraction.

We note in passing that the relative weight between quarks and gluons within the electric component, while an informed one, is nevertheless a model assumption. It will be a future project to investigate the constraining power of available experimental data on an expanded parameter space. Additionally, it would be interesting to explore other potentially relevant non-perturbative degrees of freedom.

\vspace{2mm}
\emph{Bayesian inference for \textsc{cujet} parameters.}---
BI quantifies the uncertainty of model parameters ($\theta$) through the posterior distribution $p(\theta | \mathcal{D})$. This uncertainty is then propagated to model predictions, which enables the estimation of credible intervals. For a given dataset $\mathcal{D}$, the posterior reads:
\begin{equation}
 p(\theta | \mathcal{D}) = \mathcal{N}_{\mathrm{norm}}\,  p(\mathcal{D} | \theta)\, p(\theta),
 \label{eq:bayes}
\end{equation}
where $\mathcal{N}_{\mathrm{norm}}$ is the normalization constant and $p(\theta)$ encodes prior knowledge of the model parameters. In this work, the priors are specified as follows:
\begin{enumerate}[leftmargin=.8\parindent]
\item $\alpha_C$ is assigned a uniform prior over $[0.3, 1.34]$.
\item $c_M$ follows a uniform prior in $[0.19, 0.56]$, consistent with lattice QCD constraints~\cite{Nakamura:2003pu, Shi:2018izg}.
\item $\chi_T(T)$ is discretized at 10 temperature points, $T \in \{0.16, 0.20, \cdots, 0.52\}~\mathrm{GeV}$, and linearly interpolated within \textsc{cujet}. 
We tested two prior functional forms for $\chi_T(T)$. In the first scenario, the $\chi$ values at different temperature points are treated as completely independent and uncorrelated, each being assigned a uniform prior in $[0.0, 1.0]$. In the second scenario, $\chi_T(T)$ is constrained to monotonically increase with $T$, consistent with a quite general physical scenario~\cite{Liao:2005pa,Liao:2008jg,Hidaka:2008dr,Hidaka:2009ma}. They are respectively referred to as {\bf \em ``unconstrained''} and {\bf \em ``monotonic''} throughout this paper.
\end{enumerate}

\begin{figure}[!ht]
    \centering
    \includegraphics[width=\linewidth]{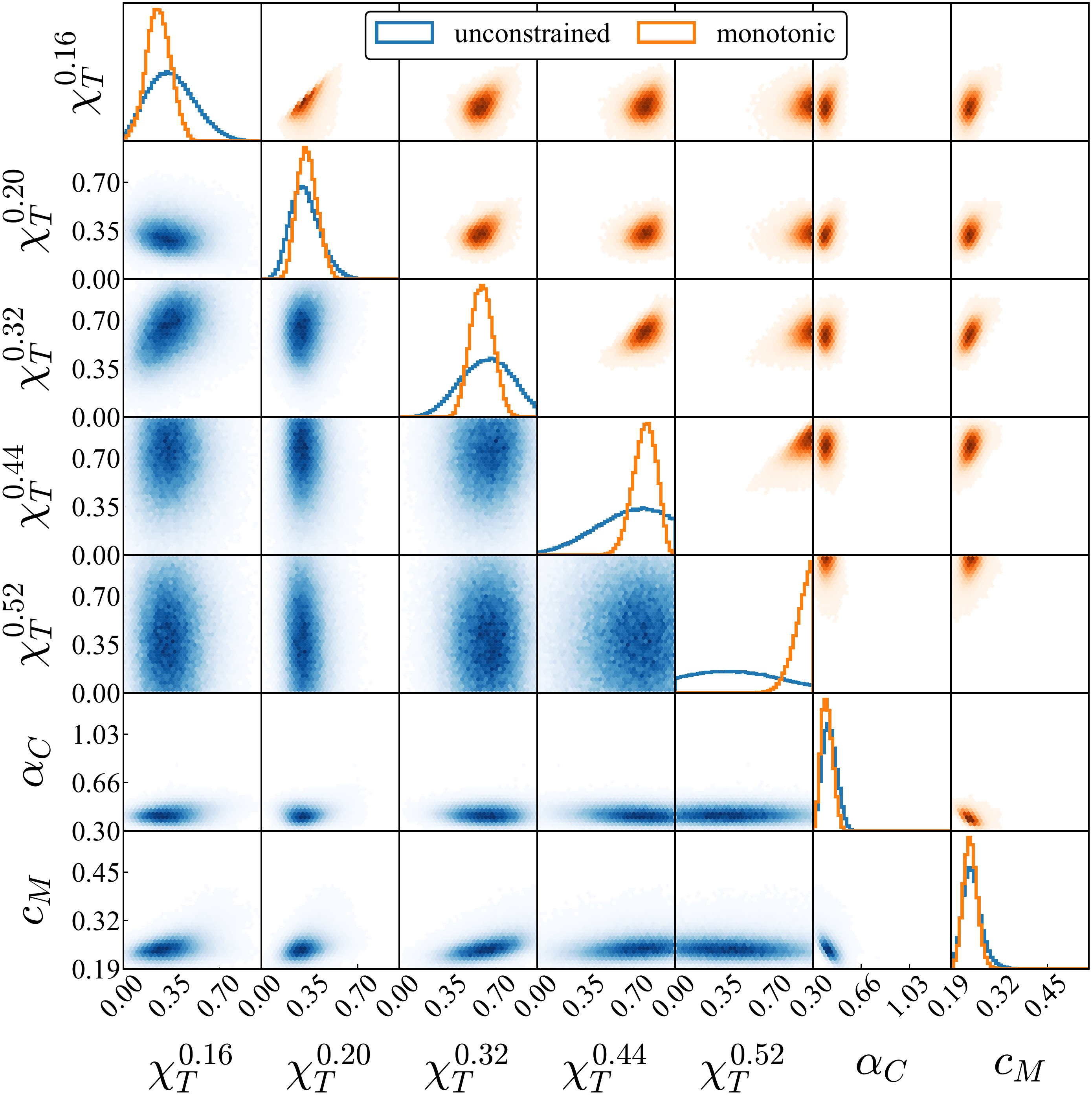}
    \caption{Posterior marginal and pairwise distributions of parameters. Blue regions (lower triangle) show samples from the unconstrained posterior, while orange regions (upper triangle) represent results from the monotonic scenario for $\chi_T(T)$.  }
    \label{fig:posteriors}
\end{figure}

In Eq.~\eqref{eq:bayes}, the likelihood $p(\mathcal{D} | \theta)$ measures how well the model predictions match the experimental observations, given the parameters $\theta$. 
We model the likelihood as a multivariate Gaussian,
\begin{align}
\log p(\mathcal{D} \mid \theta) 
= -\frac{1}{2} \Big[
&(\mathbf{y} - \hat{\mathbf{y}})^{\mathrm T} 
\Sigma^{-1} 
(\mathbf{y} - \hat{\mathbf{y}}) \nonumber \\
&+ \log \det \Sigma 
+ N \log(2\pi)
\Big],
\end{align}
where $\hat{\mathbf{y}}$ is the \textsc{cujet} prediction and $\Sigma$ is the covariance matrix. Since full experimental covariance matrices are unavailable, our primary analysis uses diagonal uncertainties, $\sigma_i^2=\sigma_{\mathrm{exp},i}^2+\sigma_{\mathrm{model},i}^2$. We also tested block-wise correlated normalization uncertainties and found stable posterior distributions.
See Refs.~\cite{Bernhard:2016tnd, Bernhard:2019bmu, JETSCAPE:2020shq, JETSCAPE:2020mzn, JETSCAPE:2021ehl, JETSCAPE:2024cqe, Heffernan:2023gye, Heffernan:2023utr, Xie:2022ght, Xie:2022fak, Gonzalez:2020bqm, MUSES:2023hyz, Domingues:2024pom, Jena:2025xcf, Grishmanovskii:2025mnc, Altenkort:2023eav, Liu:2023rfi, Jahan:2024wpj, Nijs:2020ors, Nijs:2020roc, Chen:2025fpd, Zhu:2026vve} for examples of extracting QCD thermodynamic and transport properties from relativistic heavy-ion collisions with BI.

\begin{figure}
    \centering
    \includegraphics[width=0.235\textwidth]{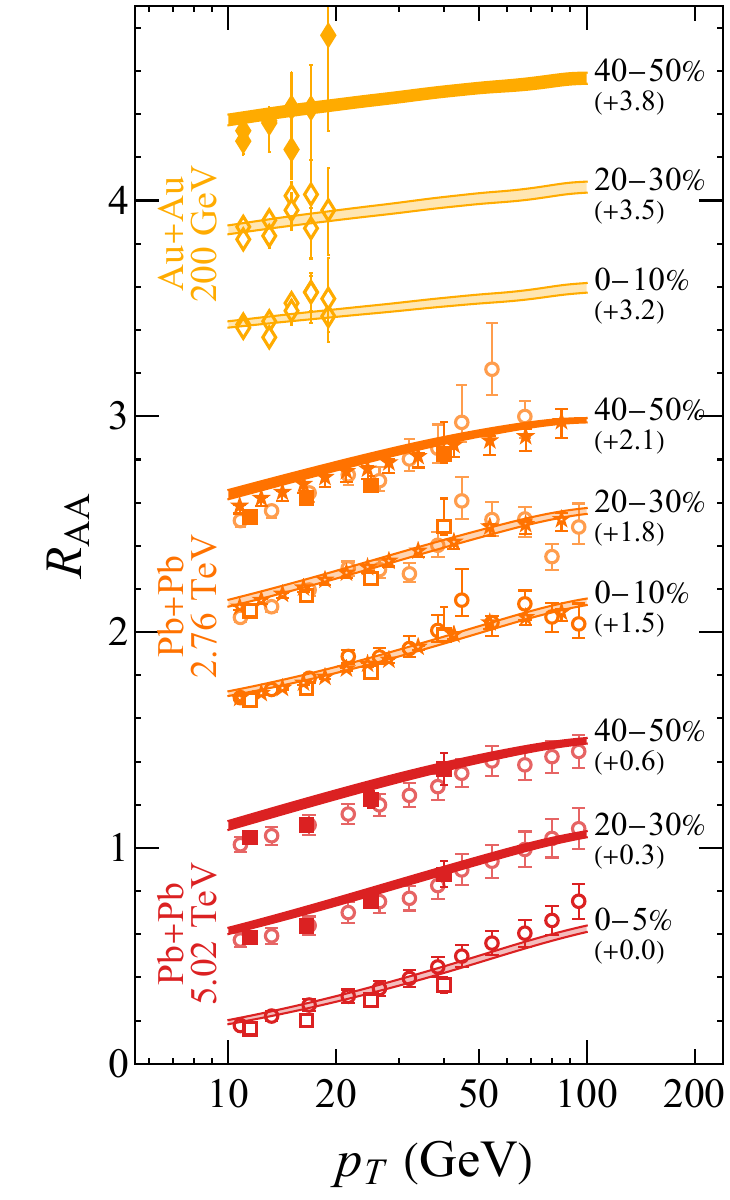}
    \includegraphics[width=0.235\textwidth]{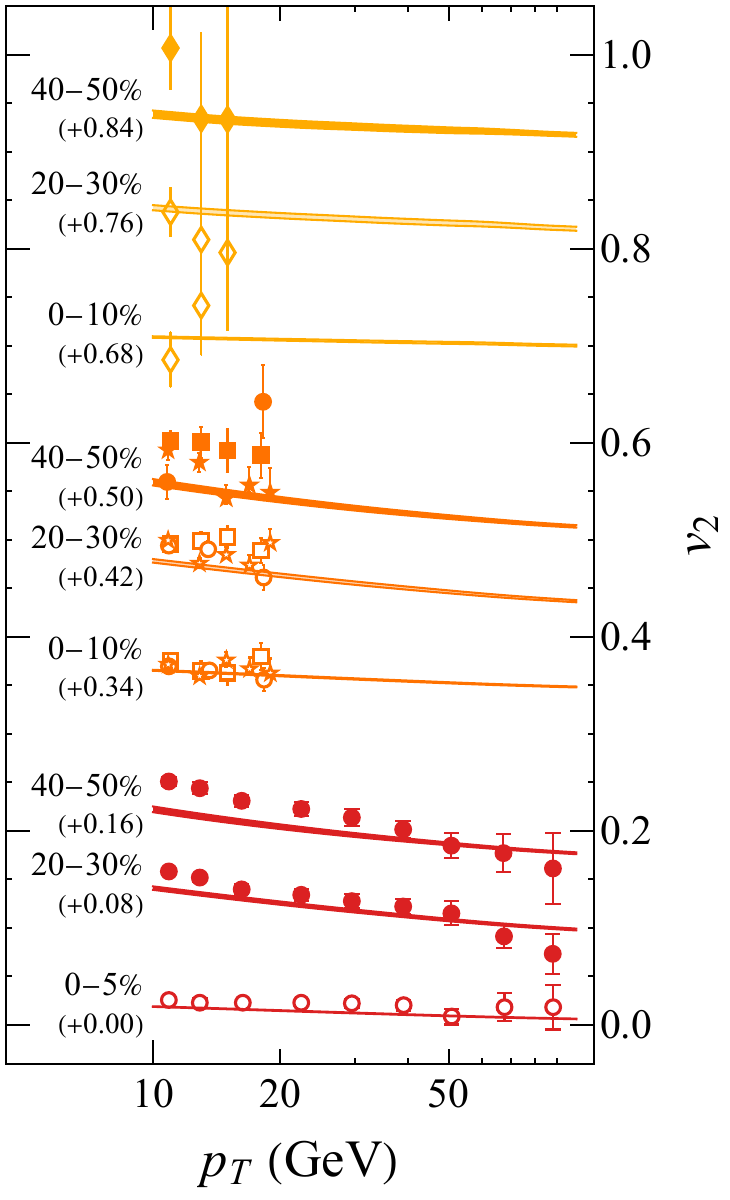}
    \caption{Posterior predictive distributions (PPDs) for $R_{AA}$ (left) and $v_2$ (right) computed from CUJET, using parameter samples located within the 95\% joint credible boundary of the constrained posterior distributions. Centrality intervals serving as validation (training) datasets are represented by filled (transparent) theory bands and solid (open) data points.
    Experimental data from PHENIX~\cite{Adare:2008qa, Adare:2012wg, PHENIX:2013yhu}, ALICE~\cite{ALICE:2018vuu, ALICE:2012vgf}, ATLAS~\cite{ATLAS:2015qmb, ATLAS:2011ah}, and CMS~\cite{CMS:2012aa, CMS:2012zex, Khachatryan:2016odn, Sirunyan:2017pan} collaborations are respectively shown as diamond, square, star, and circle symbols.
    In the $R_{AA}$ plot, open circles with lighter color are CMS data within $10-30\%$ and $30-50\%$ centrality bins. 
    \label{fig:PPDs}}
\end{figure}
Since direct evaluation of \textsc{cujet} is computationally expensive during posterior exploration, we employ a Gaussian Process Emulator (GPE) implemented in \textsc{gpytorch}~\cite{gardner2018gpytorch} as a surrogate model. 
The GPE is trained on 650 parameter points generated via Maximin Latin Hypercube Sampling (LHS) in the 12-dimensional parameter space, with an additional 100 points reserved for validation. 
The emulator is validated on an independent test set using root-mean-squared error, maximum absolute error, and predictive-interval coverage, showing good predictive accuracy and reasonably calibrated uncertainties. To avoid overconfidence, we use
\begin{equation}
\sigma_{\mathrm{model}}^2=\sigma_{\mathrm{GPE}}^2+\sigma_{\mathrm{noise}}^2,
\end{equation}
where $\sigma_{\mathrm{noise}}^2$ is estimated from the validation error to account for possible residual GPE miscalibration.
For the Bayesian analysis of this work, we utilize six experimental datasets, as summarized in the Table below. These  include a total of 100 individual data points and span  a broad range of $p_T$ and collision centrality bins. 
Posterior sampling is performed with the No-U-Turn Sampler (NUTS)~\cite{hoffman2014nuts}, a self-adaptive extension of Hamiltonian Monte Carlo, implemented via the probabilistic programming framework \textsc{pyro}~\cite{bingham2019pyro}.
\begin{table}[htbp]
    \centering
    \caption{Experimental datasets used in the analysis.}
    \label{tab:exp_data}
    \begin{tabular}{llll}
        \hline \hline
        Collab./Ref. & System & Centr. (\%) & $p_T$ (GeV/$c$) \\
        \hline
        \multirow{2}{*}{PHENIX~\cite{Adare:2012wg,PHENIX:2013yhu}}
          & Au-Au            & \multirow{2}{*}{0--10, 20--30} & $R_{AA}^{\pi^0}$: [7, 19] \\
          & 200 GeV          &                                & $v_2^{\pi^0}$:\ \ \ [5, 16] \\
        \hline
        \multirow{2}{*}{ATLAS~\cite{ATLAS:2015qmb,ATLAS:2011ah}}
          & Pb-Pb            & \multirow{2}{*}{0--10, 20--30} & $R_{AA}^{ch}$: [6.68, 59.8] \\
          & 2760 GeV         &                                & $v_2^{ch}$:\ \ \ [8, 18] \\
        \hline
        \multirow{2}{*}{CMS~\cite{Khachatryan:2016odn,Sirunyan:2017pan}}
          & Pb-Pb            & \multirow{2}{*}{0--5}          & $R_{AA}^{ch}$: [6.4, 54.4] \\
          & 5020 GeV         &                                & $v_2^{ch}$:\ \ \ [12.9, 67.7] \\
        \hline \hline
    \end{tabular}
\end{table}

Figure~\ref{fig:posteriors} shows the posterior marginal and pairwise joint distributions of key model parameters obtained from BI in both unconstrained and monotonic scenarios. (For better readability, we omit intermediate temperature points of $\chi_T(T)$ at   $T = 0.24$, $0.28$, $0.36$, $0.40$, and $0.48~\mathrm{GeV}$.) We first notice that the parameters $\alpha_C$ and $c_M$ are well constrained and consistent between both scenarios.

\begin{figure}[!hbtp]
    \centering
    \includegraphics[width=0.5\textwidth]{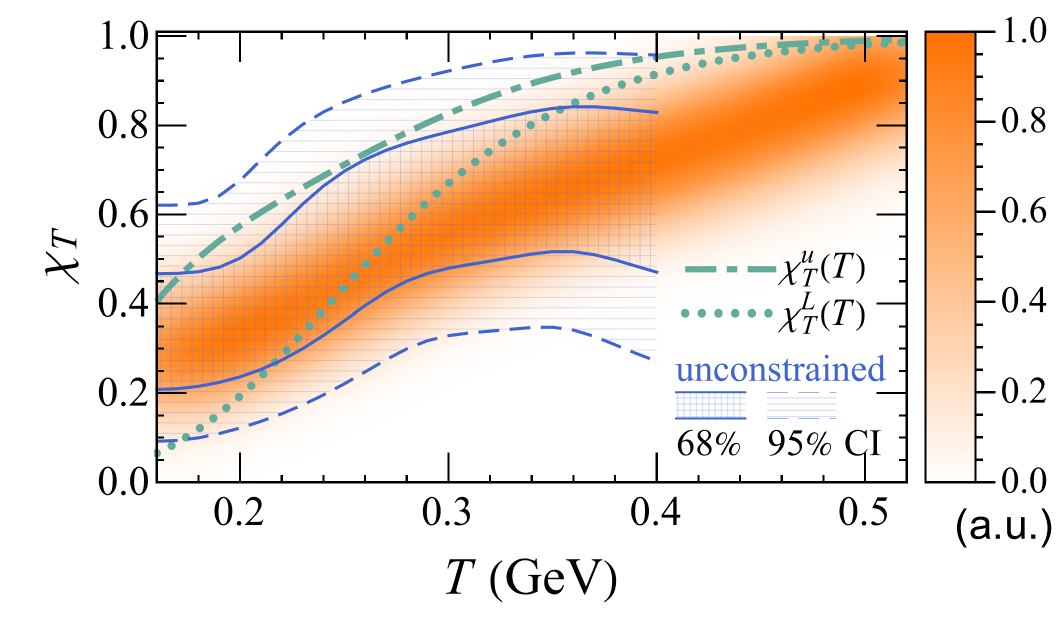}
    \includegraphics[width=0.5\textwidth]{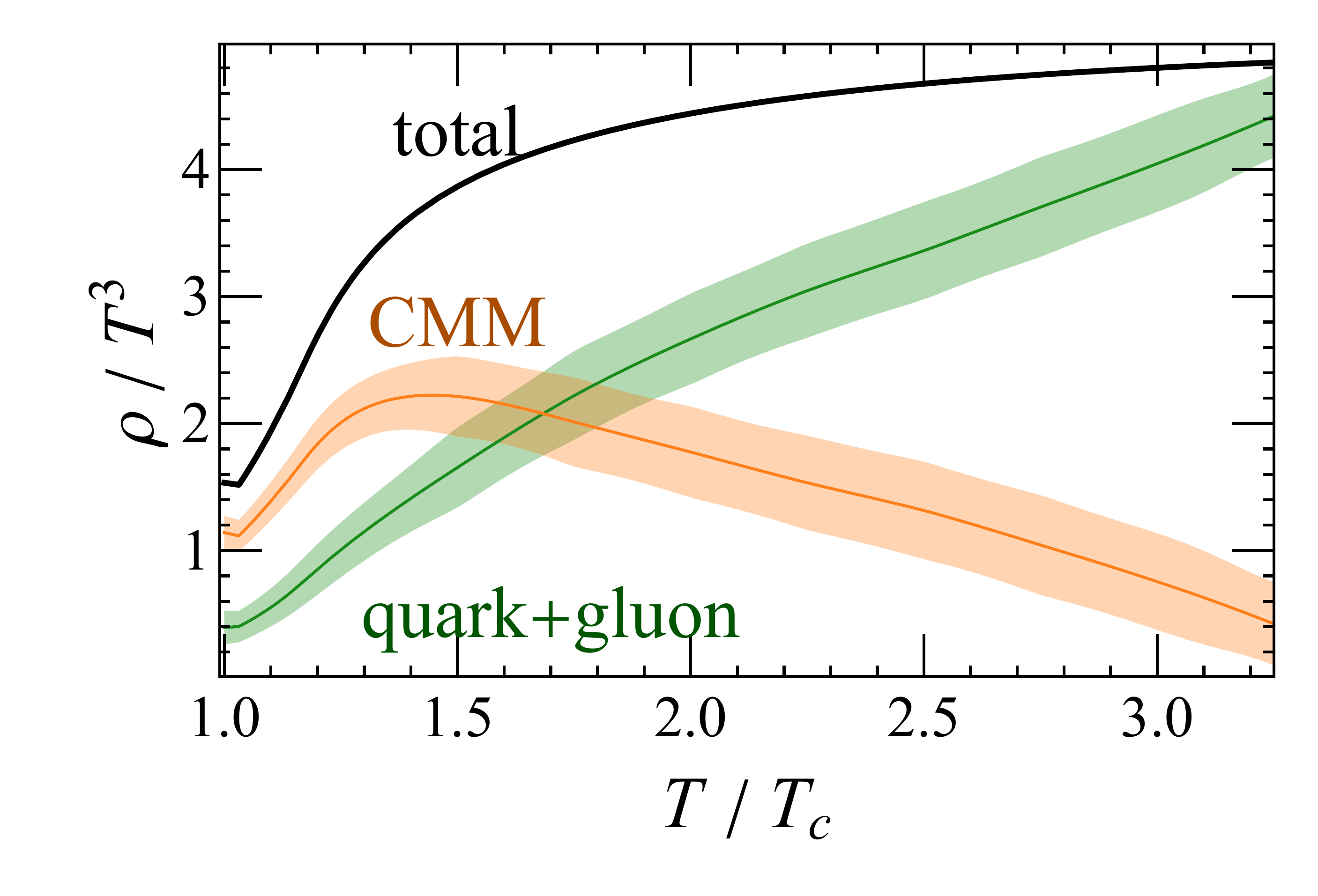}
    \caption{(Upper) Posterior distribution of $\chi_T(T)$ in the monotonic scenario is shown as orange band.  
    The $68\%$ and $95\%$ credibility intervals from Bayesian analysis in the unconstrained scenario (blue shaded bands) as well as the \textsc{cujet3.1} model assumptions of $\chi_T$ based on quark number susceptibilities ($\chi_T^u$, green dash-dotted) and Polyakov loop ($\chi_T^L$, green dotted) are shown for comparison. (Lower) The temperature dependence of the normalized  number densities $\rho/T^3$ for CMM's (orange band), for quarks and gluons (green band), as well as for all particles (black line),  that are from the BI results with monotonic scenario.
    \label{fig:chiT_T}}
\end{figure} 
Let us then focus on the posterior results for $\chi(T)$ in the {\em unconstrained scenario}, shown in the lower triangle. In the near-$T_c$ temperature region, the peaks in the  single-variable marginal distributions indicate that $\chi_T$ remains significantly below unity and increases monotonically with $T$. This behavior suggests a substantial fraction of CMMs, whose abundance decreases gradually with increasing temperature~\cite{Liao:2005pa}.
In the high-temperature region, the posterior marginal distributions for $\chi_T$ gradually widen, indicating a reduced constraining power of experimental data. Such a reduced sensitivity is simply a consequence of the fact that  such high temperatures (e.g., $T=0.44~\mathrm{GeV}$ and up) account for only a very small part of the overall space-time evolution of the QGP fireball, even for the most central collisions at $5.02~\mathrm{TeV}$.

\begin{figure*}[!ht]
    \centering
    \includegraphics[width=0.4\textwidth]{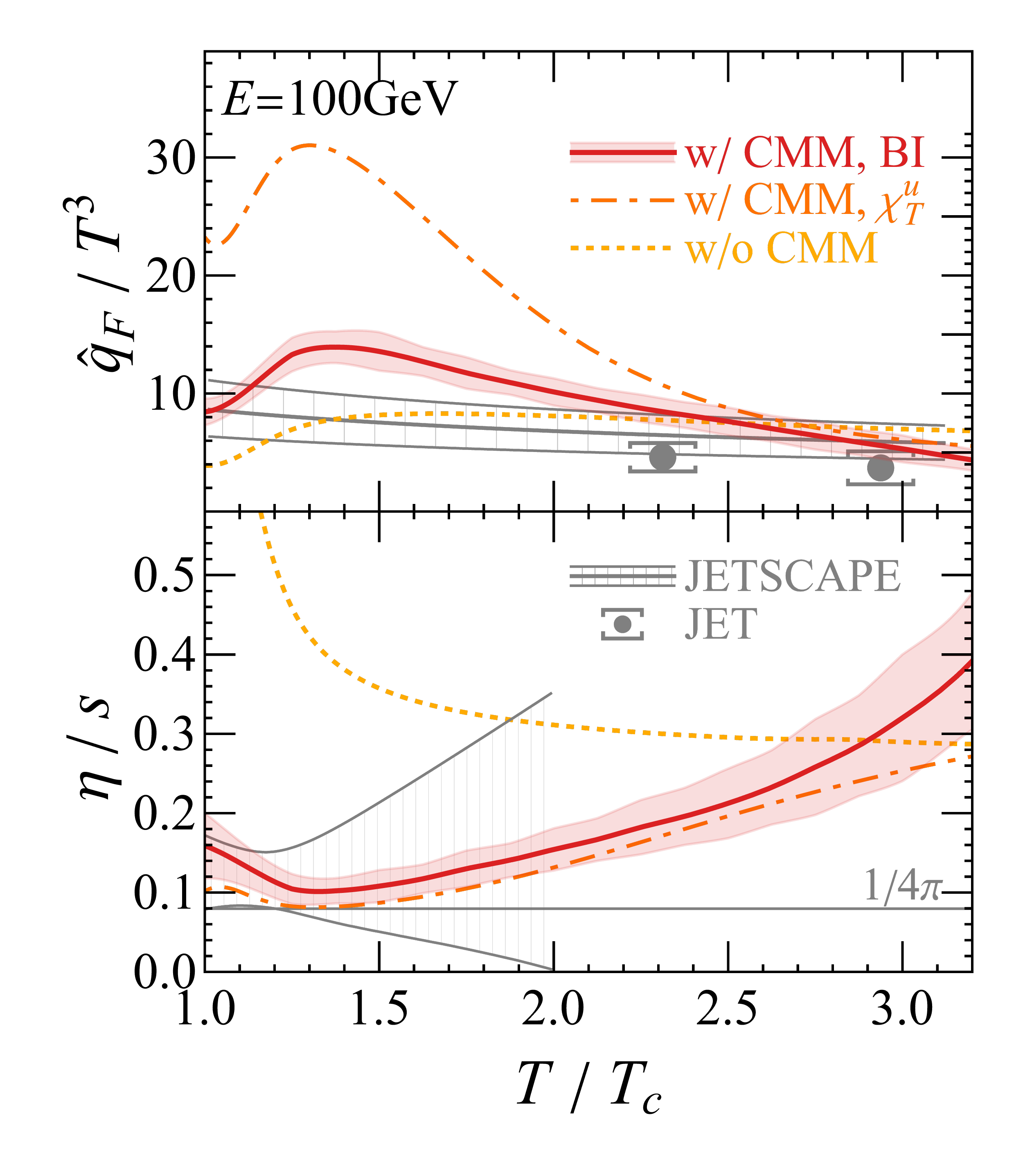}\qquad
    \includegraphics[width=0.4\textwidth]{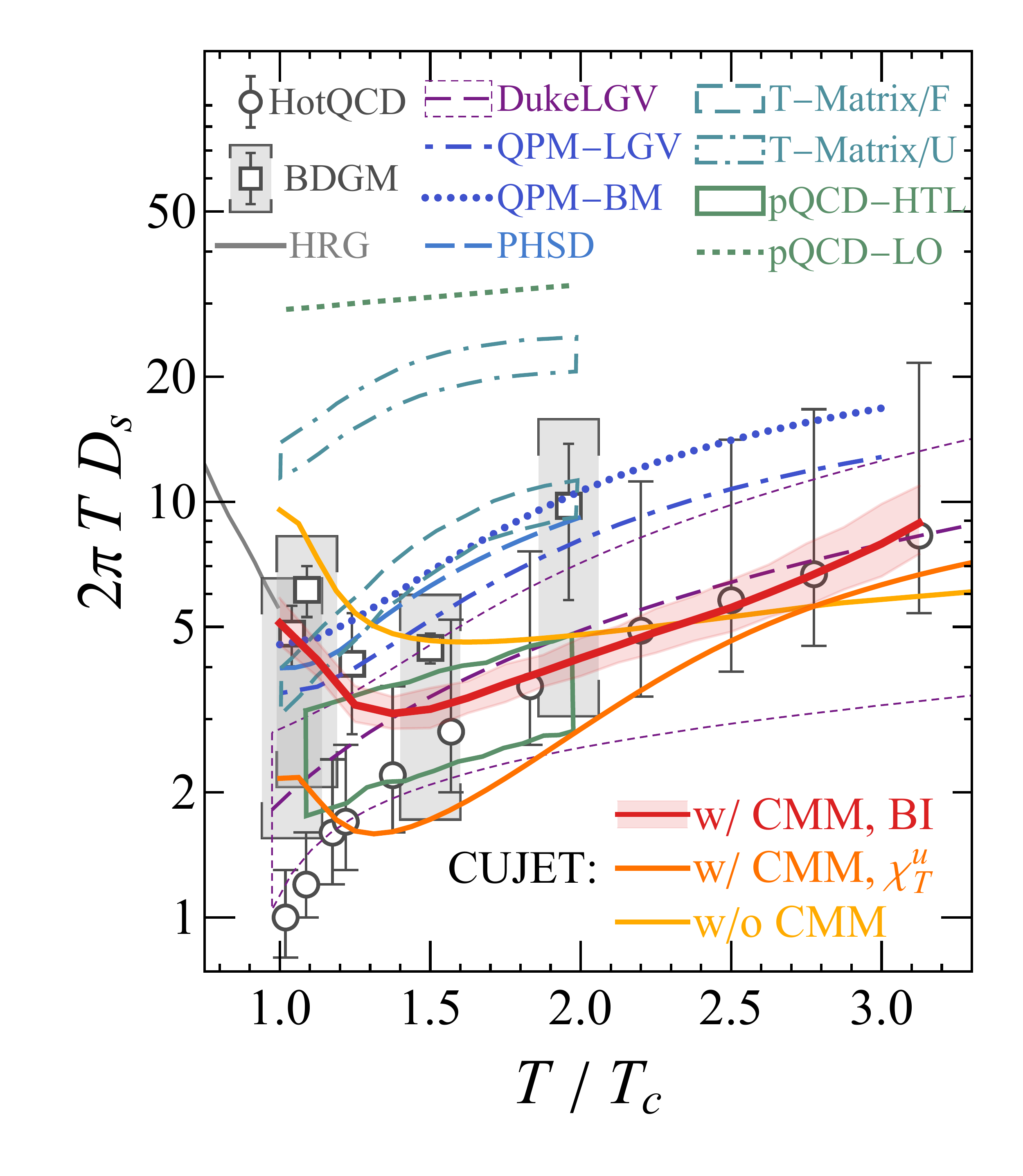}
    \caption{Transverse-momentum-transfer-squared per unit path length (upper left), shear viscosity (lower left), and diffusion parameter (right) are shown as functions of temperature. In all sub-figures, red lines and bands respectively represent the median and $95\%$ credibility interval of the current Bayesian Inference (BI); orange lines are \textsc{cujet3.1} results that estimates the chromo-magnetic degrees of freedom from quark susceptibility ($\chi_T^u$); gold lines correspond to \textsc{cujet2} simulations without CMM. Results from other models are included for comparison: \textsc{jet} collaboration model-combined extraction of $\hat{q}$~\cite{JET:2013cls}; \textsc{jetscape} Bayesian analyses of $\hat{q}$ based on energetic particles~\cite{JETSCAPE:2020shq, JETSCAPE:2020mzn} and $\eta/s$ for soft particles~\cite{JETSCAPE:2024cqe}; diffusion parameter computed from lattice QCD simulations~\cite{Banerjee:2011ra, Ding:2012sp, HotQCD:2025fbd}, \textsc{UrQMD}'s simulation based on hadron resonance gas(\textsc{hrg}) model~\cite{Lang:2012nqy}, Duke's Langevin-based model~\cite{Cao:2011et, Cao:2012jt}, Catania's quasi-particle models(\textsc{qpm}) for solving the Boltzmann(\textsc{bm}) and Langevin(\textsc{lgv}) equations~\cite{Scardina:2017ipo}, \textsc{phsd}~\cite{Bratkovskaya:2011wp, Song:2015sfa, Song:2015ykw}, TAMU's T-Matrix calculation using free-energy(F) and inner-energy(U) assumptions~\cite{vanHees:2007me, Riek:2010fk, Liu:2016ysz}, and \textsc{amy} hard thermal loop(\textsc{htl})~\cite{Arnold:2003zc} and SUBATECH's leading order(\textsc{lo})~\cite{Gossiaux:2008jv, Peshier:2008bg} perturbative QCD calculations.}
    \label{fig:transport_coefficients_mono}
\end{figure*}
The uncertainty of $\chi_T$ at high temperatures can be naturally resolved once  we require  that $\chi_T$  increases  monotonically with $T$ — a physical constraint that is well motivated by UV asymptotic freedom and IR  electric-magnetic duality for non-Abelian gauge theories. 
The upper triangle  of Fig.~\ref{fig:posteriors} shows the resulting posterior distributions in this {\em monotonic scenario}. 
As one can see,  the $\chi_T$ distributions at all temperatures become much narrower, leading to a statistically strong determination of $\chi_T$ values across temperatures.

As an independent verification, we compare the monotonic CMM model (denoted as $M_1$) with a no-CMM baseline (labeled $M_0$), where the CMM component is switched off and only $(\alpha_C,c_M)$ remain free. With $B_{10}=Z_1/Z_0$ and $Z_i=\int P(\mathcal{D}|\theta_i,M_i)P(\theta_i|M_i)d\theta_i$, we find $\log B_{10}\simeq4.99$ ($B_{10}\simeq1.47\times10^2$), providing strong evidence that the data favor $M_1$ over $M_0$ within the \textsc{cujet} framework.

Next, we extract parameter combinations along the boundary of the 95\% joint credible region and use them as conservative inputs for the \textsc{cujet} calculations, thereby generating the Posterior Predictive Distributions (PPDs). These PPDs are used both to verify the convergence of the BI and to extend the theoretical predictions to a wider range of experimental observables beyond those used for BI. Figure~\ref{fig:PPDs} shows the resulting $R_{AA}$ and $v_2$ predictions from the monotonic scenario, and we have checked that the unconstrained scenario results in quantitatively consistent theory predictions. As one can see, the PPDs provide an overall reasonable description of the $R_{AA}$ and $v_2$ data across different collision energies and centrality classes. Noticeable discrepancies remain for the $v_2$ measurements in the 20--30\% and 40--50\% Pb–Pb collisions at 2.76 TeV, particularly toward the lower-$p_T$ region. Such deviations may indicate contributions from additional mechanisms not included in the present CUJET implementation. For example, hadronization mechanisms such as coalescence or transverse momentum broadening effect may contribute to the measured anisotropy toward lower $p_T$, beyond the fragmentation-based description adopted here. 
It may be noted that due to the highly nonlinear nature of the jet energy loss phenomenon, it is not possible to clearly separate the electric and magnetic contributions in the final state observables $R_{AA}$ and $v_2$. Nevertheless, within the CUJET framework, the Bayesian analysis shows a strong preference for a substantial magnetic component, indicating that near-$T_c$ enhancement of jet-medium interactions plays an important role in describing these observables.

Finally, let us focus on comparing  $\chi_T$ at different temperatures from BI with and without monotonic constraints. In Fig.~\ref{fig:chiT_T} (upper) we present the posterior distribution of $\chi_T(T)$ in the monotonic scenario (orange band),  along  with the $68\%$ and $95\%$ credible intervals in the unconstrained scenario (blue shaded bands). Interestingly, we note that even in the unconstrained scenario, the $\chi_T$  increases monotonically in the region $T_c \leq T \lesssim 0.35~\mathrm{GeV}$ and aligns well with $\chi_T$ in the monotonic scenario, which implies that the monotonic trend is strongly supported by experimental data. This trend in the unconstrained scenario becomes unclear when entering the higher temperature region, which simply indicates the reduced constraining power of data alone for those temperatures. This is exactly where a well-justified physical consideration, i.e. a monotonic constraint, becomes useful. Therefore, we adopt the monotonic scenario in all subsequent discussions. It is also interesting to compare the $\chi_T(T)$ from the current BI with previous \textsc{cujet3.1} model assumptions, also shown as dash-dotted and dotted curves in Fig.~\ref{fig:chiT_T} (upper).   
For temperatures just above $T_c$, the $\chi_T$ from BI lies between the susceptibility-based~\cite{McLerran:1987pz, Gottlieb:1988cq, Gavai:1989ce, Gottlieb:1987ac} ($\chi_T^u$) and Polyakov-loop-based~\cite{Hidaka:2008dr, Hidaka:2009ma, Dumitru:2010mj, Lin:2013efa} ($\chi_T^L$) approximations. At even higher temperatures, the posterior distribution for $\chi_T(T)$ suggests a considerably higher abundance of CMMs than $\chi_T^u$ or $\chi_T^L$. As the CMMs cause more jet energy loss, the present BI obtains  a very reasonable value of the  posterior coupling parameter  ($\alpha_C \approx 0.4$),  as compared to the much larger value ($\alpha_C \approx 0.9$) obtained from $\chi^2$-fitting analysis in the previous \textsc{cujet3.1} model based on $\chi_T^u$ or $\chi_T^L$.

With the BI estimated $\chi_T(T)$, the corresponding number densities of CMMs as well as quarks and gluons are shown in Fig.~\ref{fig:chiT_T}. 
The CMM density rapidly increases from $T_c$ and peaks around $T\sim1.5\,T_c$, becoming the dominant degree of freedom in this temperature region. Only at  higher temperatures $T \gtrsim 2T_c$, the quarks and gluons gradually take over and dominate the QGP. These results transform the qualitative scenario of Fig.~\ref{fig:illustration} into a quantitative description of the microscopic structures of the quark-gluon plasma.

\vspace{2mm}
\emph{Hard and soft transport coefficients}.
With Bayesian extraction of \textsc{cujet} parameters including $\chi_T$ as well as $\alpha_C$ and $c_M$, we thus arrive at a data-driven determination of the strongly coupled QGP with specific compositions of chromo-electric and chromo-magnetic quasiparticles. It is of great interest to compute key transport properties of such a system. Let us start with the so-called jet transport coefficient $\hat{q}$ that quantifies the average transverse-momentum-transfer-squared per unit path length: 
\begin{align}
\hat{q}_a(E,T)=
    \int_0^{6ET}\mathbf{q}_\perp^2 d^2\mathbf{q}_\perp
    \sum_{b} C_{ab}\, \rho_b\,\alpha_s^{i_b}(\mathbf{q}_\perp^2) \frac{\mathrm{d}\sigma_{b}}{\mathrm{d}\mathbf{q}_\perp^2},
\end{align}
in which $a,b \in \{g,q,m\}$, $i_b=0$ for quarks and gluons, and $i_b=-2$ for CMMs. The color factor $C_{ab}$ takes the value of $4/9$ for quark-quark scattering, $9/4$ for gluon/CMM-gluon/CMM, and $1$ otherwise. We take the commonly adopted labeling $\hat{q}_F \equiv \hat{q}_q$ later on.
The above $\hat{q}$ can be extrapolated down to the soft energy scale and used for estimates of bulk transport coefficients based on the kinetic transport scenario~\cite{Danielewicz:1984ww, Hirano:2005wx, Majumder:2007zh,Xu:2014tda,Xu:2015bbz,Shi:2018lsf,Shi:2018izg}. This   gives the shear-viscosity-to-entropy-density ratio as
\begin{align}
    \frac{\eta}{s} = \frac{18T^3}{5s} \sum_{a \in \{g,q,m\}} \rho_a/\hat{q}_a(E=3T/2),
\end{align}
and the heavy-quark diffusion parameter as
\begin{align}
    D_s = 4T^2/\hat{q}_F(E=3~\mathrm{GeV}).
\end{align}
The $\hat{q}_F/T^3$, $\eta/s$, and $2\pi T D_s$ based on our BI results are presented in Fig.~\ref{fig:transport_coefficients_mono}. Also included for comparison are corresponding results from lattice QCD calculations~\cite{Banerjee:2011ra, Ding:2012sp, HotQCD:2025fbd}, other BIs~\cite{JETSCAPE:2024cqe, JETSCAPE:2020mzn, JETSCAPE:2020shq}, or other model calculations~\cite{JET:2013cls, Kovtun:2004de, Cao:2011et, Cao:2012jt, Scardina:2017ipo, Bratkovskaya:2011wp, Song:2015sfa, Song:2015ykw, vanHees:2007me, Lang:2012nqy, Riek:2010fk, Liu:2016ysz, Arnold:2003zc, Gossiaux:2008jv, Peshier:2008bg}. See e.g.~\cite{Rapp:2018qla} for a comprehensive review. Results from two previous \textsc{cujet} model analyses,  the \textsc{cujet3.1}~\cite{Shi:2018izg} with CMM's fraction estimated from quark susceptibility ($\chi_T^u$) and the \textsc{cujet2}~\cite{Xu:2014ica} without CMM, are also shown. 

We observe that  the current BI result of $\hat{q}_F/T^3$  is in broad agreement with the joint-model comparison from the \textsc{jet} collaboration~\cite{JET:2013cls}, as well as the \textsc{jetscape} collaboration's BI of $\hat{q}$ assuming a \textsc{matter}+\textsc{lbt} jet energy-loss mechanism~\cite{JETSCAPE:2024cqe}, but our temperature dependence is more pronounced due to the presence of the chromo-magnetic component and its $T$-dependence via $\chi_T$. By comparing the \textsc{cujet} results with and without chromo-magnetic components, we observe a clear CMM-induced enhancement of jet energy loss peaked around $T \sim 1.3\,T_c$ where the CMMs are dominant.

The chromo-magnetic degrees of freedom also provide important insights into understanding the shear viscosity of QCD matter. 
Whereas the case without CMM (\textsc{cujet2}) leads to a large value of $\eta/s \gtrsim 0.3$ in the relevant temperature region, the scenarios with CMMs --- including both BI extraction and quark-susceptibility estimation --- exhibit considerably  smaller $\eta/s$ values that are needed for  hydrodynamic simulations to describe an abundance of collective flow observables for low-$p_T$ particles~\cite{JETSCAPE:2020mzn, JETSCAPE:2020shq}. It can be seen that $\eta/s$ reaches a minimum value slightly above the KSS lower bound of $1/(4\pi)$~\cite{Kovtun:2004de} in the region close to $T_c$. 
This comparison clearly supports the conclusion that emergent CMMs play a key role in explaining the nearly perfect fluidity of QGP near $T_c$. 

Lastly, the heavy-quark diffusion parameter extracted from the current BI is also in good agreement with lattice QCD calculations~\cite{Banerjee:2011ra, Ding:2012sp, HotQCD:2025fbd} as well as with various other model estimations, further validating a QGP with a significant magnetic component especially close to $T_c$. Notably, with the BI extracted values for CMM fraction and strong coupling constant, the computed diffusion parameter matches well with the value below $T_c$ computed from scatterings in a  hadron resonance gas~\cite{Lang:2012nqy}. It shows that  the chromo-magnetic degrees of freedom naturally connect the hadronic phase below $T_c$ with the high-temperature phase of quarks and gluons.

\vspace{2mm}
\emph{Summary.}---In this work, we have performed the first Bayesian analysis to determine the temperature-dependent fraction of the chromo-magnetic component in the quark-gluon plasma, based on the  \textsc{cujet} framework and utilizing a wealth of experimental data on the $p_T$-dependent $R_{\mathrm{AA}}$ and $v_2$ of high $p_T$ hadrons. The obtained posterior results show that, within the CUJET framework, the experimental data furnish a quantitative extraction of the temperature-dependent abundance of the CMMs and strongly favor a substantial  chromo-magnetic component, particularly in the near-$T_c$ region.  The resulting posterior predictions provide an overall good description of high-$p_T$ hadron $R_{AA}$ and $v_2$ across collision energies and centralities, although discrepancies remain for peripheral Pb--Pb $v_2$ toward lower $p_T$ which may imply possible contributions from mechanisms beyond those included in the present framework. Additionally, they have been shown to quantitatively describe key transport properties of QGP, such as the jet transport coefficient, the shear-viscosity-to-entropy-density ratio, as well as the heavy-quark diffusion constant. All taken together, we conclude that the present study provides by far the strongest data-driven evidence for the important role of emergent chromo-magnetic monopoles in the quark-gluon plasma.

\vspace{2mm}
\textbf{Acknowledgments.} --- The authors are grateful to Dr. Raymond Ehlers and Dr. Jean-Francois Paquet for kindly sharing the JETSCAPE results. We also thank Shuhan Zheng for helpful discussions. JL and SS thank Dr. Miklos Gyulassy and Dr. Jiechen Xu for previous collaborations that developed the CUJET3 framework. 
YG and SS acknowledge support from the National Key Research and
Development Program of China under Contract No. 2024YFA1610700. JL is supported by the U.S. NSF under Grant No. PHY-2514992. 
\bibliography{ref}
\end{document}